\documentstyle[preprint,epsfig]{jpsj}

\def\runtitle{Dynamical stability of the crack front line}
\def\runauthor{Fukuhara and Nakanishi}

\title{
Dynamical stability of the crack front line
}

\author{Takayasu {\sc Fukuhara\footnote{Present address:
Mikuniya Corp. 1-1-20, Toranomon-Jitsugyokaikan Bld.
Toranomon, Minato-ku, Tokyo 105, Japan,
}}
 and Hiizu {\sc Nakanishi}}
\inst{ Department of Physics, Kyushu University 33,
Fukuoka 812-8581}

\recdate{Jun. 1, 1998}

\abst{
Dynamical stability of the crack front line that propagates between
two plates is studied numerically using the simple two-dimensional
mass-spring model.  It is demonstrated that the straight
front line is unstable for low speed while it becomes stable for high
speed.
For the uniform model, the roughness exponent in the slower speed
region is fairly constant around 0.4 and there seems to be a
rough-smooth transition at a certain speed.
For the inhomogeneous case with quenched randomness, the transition is 
gradual.
}

\kword{
crack propagation, dynamical instability, crack front roughening,
rough-smooth transition
}

\begin{document}
\sloppy
\maketitle

\newpage

\section{Introduction}

It has been known for a long time that the straight and smooth crack
propagation becomes unstable under certain circumstances.
The ``mirror--mist--hackle'' transition of the crack surface of glass
rod\cite{Lawn93} is one of the manifestation of the instability
that the crack surface becomes rougher as the crack speed increases.
A much more controlled experiment that is related to the instability
has been done on the fracture propagation in the PMMA glass
plate,\cite{FGMS91} and the smooth crack propagation was found to
become unstable beyond a certain crack speed around 60\% of Rayleigh
speed.

On the other hand, there could be another instability in other
direction, namely, ``in-plane'' crack front instability.
In the three dimensional crack propagation, the crack front line
within the plane of crack surface could be rough under certain
circumstances.
The experiment to observe the crack front roughness was done first by
Daguier {\it et al.}\cite{DBL95} on very slow crack in metal alloys,
and they found the in-plane crack front roughness exponent to be
$0.51\sim 0.64$.

Schmittbuhl and M{\aa}l{\o}y\cite{SM97} observed the time development of
the in-plane crack front that propagates between the two Plexiglas
plates annealed together.
They found the crack front shows self-affine rough structure with the
exponent $0.55\pm 0.05$ for the crack speed much slower than the sound 
speed; $5\times 10^{-5}m/s$.

There are some theoretical approaches to the above instability:
Schmittbuhl {\it et al.}\cite{SRVM95} studied the quasi-static
dynamics driven by the stress intensity factor along the crack front
under the existence of quenched randomness.  They found from numerical
simulations the roughness exponent $\alpha=0.35$ and the dynamic
exponent $z=1.5$; these exponents are quite different from those for
the local dynamics, and the difference was attributed to the long
range interaction through stress field.

Ramanathan {\it et al.}\cite{REF97} studied the roughness exponents
for the surface under the similar situations, and found the surface
roughness exponent is $1/2$ for mode III loading, but the surface is
only logarithmically rough for mode I due to the long range
interaction.

In the present paper, we study the in-plane crack front stability
using the two dimensional mass-spring model when the crack speed is
comparable with the sound speed; the situation is quite different from
most of other experimental and theoretical works, where the
quasi-static case was investigated.

Organization of the paper is as follows:
The model is defined in \S 2, and the uniform propagating solution is
described in \S 3.  The method and results of the numerical
simulations for the two version of the model are given in \S 4, and
the summary and discussion in \S 5.

%
%
%
%
%


\section{Model}

\begin{figure}[htbp]
\begin{center}
  \leavevmode
  \psfig{file=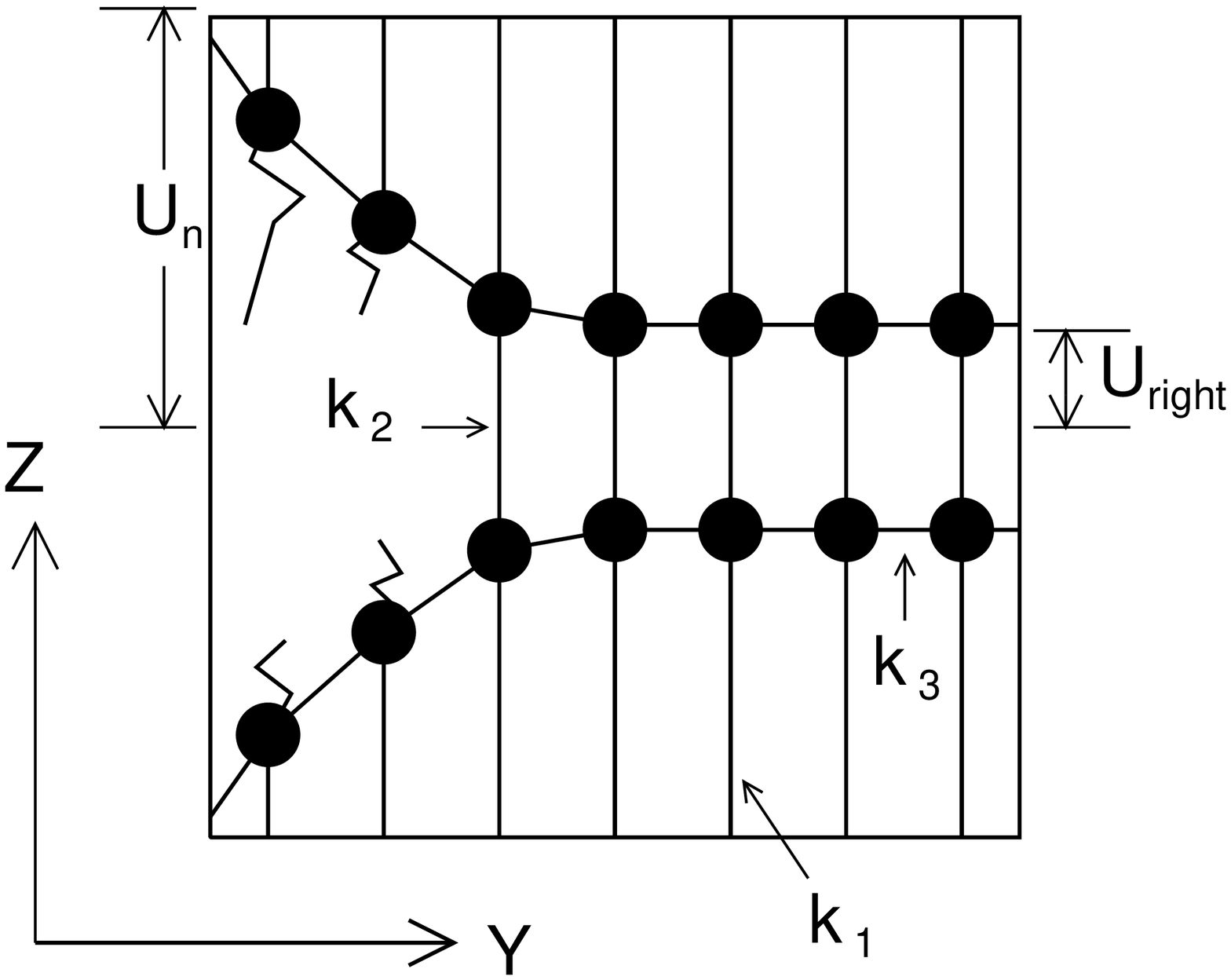,height=6cm,width=6cm} \hspace{-0.8cm}
  \psfig{file=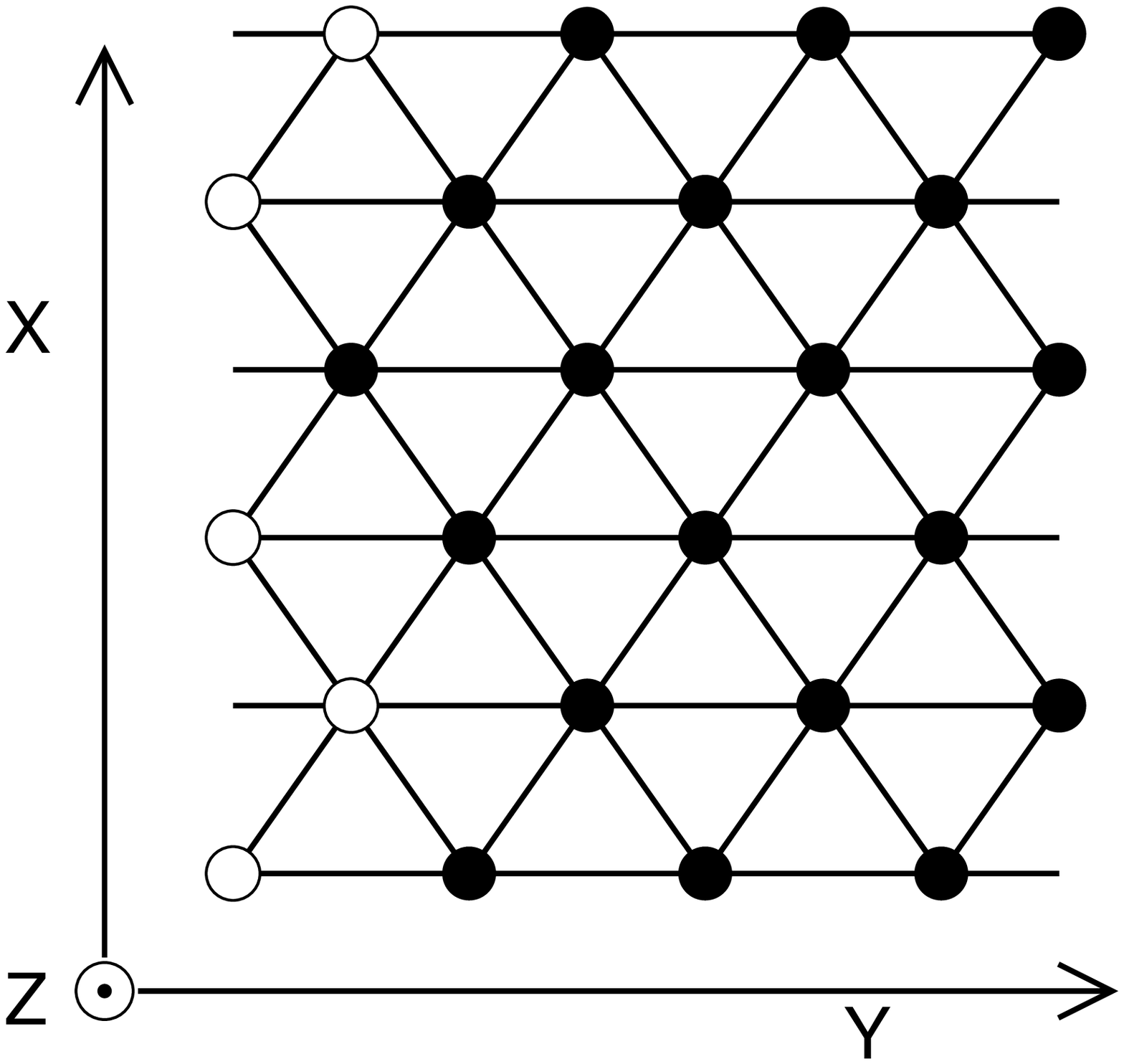,height=6cm,width=6cm}
\end{center}
\caption{
Side(left) and top(right) view of the mass-spring system.
In the top view, the open circles denote the broken bonds.
Note that we assumed the linear force with the displacement difference
for the spring $k_3$ also in eq.(1).
}
\label{fig1}
\end{figure}

The model we study is a 2-d version of the 1-d model studied by Marder
and Gross\cite{MG95}, but it is extended in the different sense from
the 2-d model by Marder and Liu,\cite{ML93} namely, the system
consists of the two layers of triangular lattices and the two plates
are attached to the system from the top and the bottom
(Fig.\ref{fig1}).  The external stress is imposed by the displacement
$2U_N$ between the two plates and the crack is assumed to propagate
along the $y$ axis by breaking the springs that connects the two
layers.

We simplify the model by assuming that the masses move only along the
$z$-axis, and the displacement of the $i$-th mass in the upper
layer is denoted by $U_i$.
Assuming the symmetric motion for the upper and lower layer, we simulate
only the upper part of the system; the displacements of the lower
masses are $-U_i$.
Then, the equation of motion we study is given by
\begin{eqnarray}
m\ddot U_i & = & k_1 (U_n - U_i) + k_2 U_i\, \theta(U_{{\rm f}i}-U_i)
\nonumber \\
& & \qquad
+ \sum_{j\in n.n.} k_3(U_j-U_i) - b \dot U_i
\label{eq.o.motion}
\end{eqnarray}
for the mass in the upper layer.
Note the factor 2 difference in the definition of $k_2$ from the one
in ref \citen{MG95}).

In the equation, $m$ is the mass, and $k_1$, $k_2$, and $k_3$ are the
spring constants that connect the $i$-th mass to the external plate,
to the mass in the other layer, and to the nearest neighbor mass in the
same layer, respectively.
(Note that we assumed the linear force with the displacement difference
for the spring $k_3$ in eq.(\ref{eq.o.motion}) although Fig.\ref{fig1}
may seem to imply the third order dependence.)
$U_{N}$ is the external displacement imposed by the plates.
$\theta(x)$ is the step function and $U_{{\rm f}i}$ is the threshold
displacement for breaking the $i$-th springs connecting the two
layers.
$b$ is the small parameter for dissipation, which is introduced to
dump the unwanted persistent oscillatory motion.

Note that the model is uniform except for the threshold $U_{{\rm f}i}$
for the spring $k_2$.


\section{Uniform propagation}

The analytic solution for the steady propagation in the 1-d version of 
the model has been obtained by Marder and Gross.\cite{MG95}
They calculated the crack speed $v$ as a function of the external
displacement and pointed out some of its interesting features:
The crack is trapped by the lattice and does not move even when the
external displacement $U_n$ exceeds the threshold value $U_n^c$ for
which the elastic energy store in the springs equals the energy 
needed to break the springs, namely, Griffith threshold, until
$\Delta\equiv U_n/U_n^c$ reaches a certain value larger than 1; then the
crack speed jumps around 0.4 time the sound speed, therefore, there is 
a forbidden band of crack speed. (Fig.5. of ref \citen{MG95}))

In the present version of the 2-d model with the uniform threshold,
the steady crack propagation shows the similar behavior: a lattice
trap and a forbidden band of crack speed.  Note that the uniform
solution in the 2-d model need not behave exactly in the same way with
the 1-d model unless we consider the crack propagation along the axis
in the square lattice, because there is a diagonal interaction between
neighboring rows.

In principle, it should be possible to do the similar analysis for the
2-d uniform solution as that of Marder and Gross\cite{MG95} by
assuming a certain periodicity of the solution, but we examine the
uniform solution numerically because our main object
of the present work is studying inhomogeneous behavior of the crack
propagation.

\begin{figure}[htbp]
\begin{center}
  \leavevmode
  \psfig{file=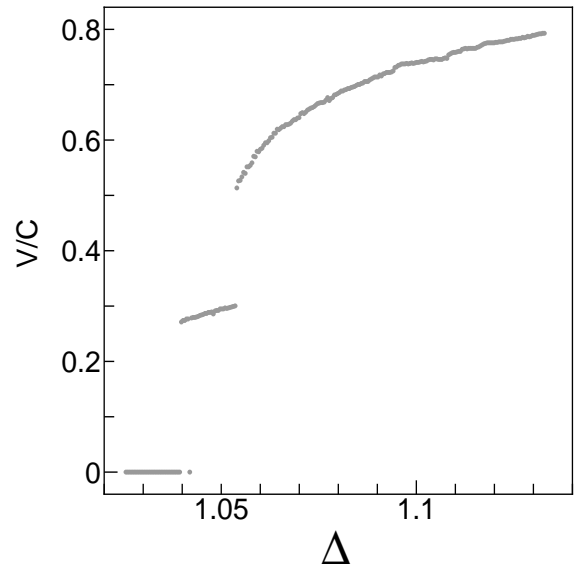,height=8cm,width=8cm}
\end{center}
\caption{
Crack speed $v$ {\it v.s.} external displacement $\Delta$ for the uniform
propagation.
$c$ denotes the sound speed.
}
\label{fig2}
\end{figure}

The numerically obtained crack speed $v$ as a function of the
normalized external displacement is shown in Fig.\ref{fig2}.  The major
difference from the 1-d case is that there is a ``mid-gap band'' in
the middle of forbidden band in the $v-\Delta$ plot. This should
come from the interaction between rows; a broken bond
helps to break the bond at the crack front in a neighboring row in the
parameter region where the 1-d model does not have a propagating
solution.


\section{Simulations and Results}

In order to examine the dynamical stability of the fracture front and
effects of inhomogeneity on it, we study two versions of the models.
The first one, which we call the model A, is the model with uniform
threshold $U_{{\rm f}i}=U_{\rm f}$ for all the site $i$
except for a single row
parallel to the crack front.
The row with random threshold is introduced to apply perturbation on
the steady solution, and we examine how the perturbation will develop 
as the crack front propagates in the uniform region.
The second one, the model B, is the model with random distribution of
the threshold over the whole system.

We simulate the system behavior by integrating eq.(\ref{eq.o.motion})
numerically using the modified Euler's method; note that the modified
Euler's method itself is the method with the second order accuracy in
the time step, but the error in linear with the time step can be
introduced by the singularity of the term that represents breaking
bonds.

We employ the unit where $m=k_2=1$,
and in all of the simulations we take $k_3=k_2$,
$k_1=0.1$, and $b=0.01$.
It is convenient to measure the external displacement by the ratio
$\Delta$,
\begin{equation}
\Delta \equiv U_n/U_n^c\, ;
\label{Delta}
\end{equation}
with $U_n^c$ being a critical displacement where the elastic energy
equals to the bond breaking energy;
\begin{equation}
U_n^c \equiv {U_n \over \sqrt{1+k_2/k_1}\, U_{\rm f}}.
\label{Unc}
\end{equation}

The boundary conditions we employ along the $x$-axis are periodic for
the model A and ``reflecting'' for the model B; in the reflecting
boundary condition, outside the system we add a virtual column which
behaves in the same way as the column next to the boundary does.
Along the $y$-axis,
only the part of the system around the crack line is
simulated, and new rows in front are added and rows left behind are
discarded as the crack advances in order that the crack front is kept
around the center of the simulated part; the length of the
simulated part is taken as about twice of the width.

\subsection{Model A}

Fig.\ref{fig3} shows the time development of the crack front
for the model with the uniform threshold $U_{{\rm f}i}=0.7$
except for a single impurity with $U_{{\rm f}i}=1.0$ located
in the 70'th row from the left side of the system
where the crack starts.
The external driving is $\Delta=1.049$.
It can be seen that the impurity causes disturbance which eventually
develops over the whole system and results in the rough crack front.

In order to avoid distortion of the crack front, next we consider the
perturbation by the impurity sites distributed over the 70'th row
randomly;
along the 70'th row, the value of threshold takes $U_{{\rm f}i}=0.8$
with the probability 0.15, and $U_{{\rm f}i}=0.7$
for the rest of the site.
The results are shown in Fig.\ref{fig4} for $\Delta=1.049$, 1.052,
and 1.053.

As the crack speed increases,
it looks the ``roughness'' of the crack front decreases.
The roughness exponents $\alpha$ of the crack front for
various values of $\Delta$ are plotted in Fig.\ref{fig5}.  We used the
Max-Min method\cite{SVR95} to determine the roughness exponents, but
we have also checked for some points that the power spectrum method
gave the similar values.  The roughness exponent $\alpha$ for the
``slow'' crack is fairly constant and about 0.4, but the crack front
becomes smooth and $\alpha\sim 0$ for $\Delta > 1.052$, which roughly
coincides with the end of the mid-gap band in Fig.2.


\subsection{Model B}

In the model B, the impurity site with $U_{{\rm f}i}=1.0$ is distributed
over the whole system with the probability 0.09 and the rest of the
sites has $U_{{\rm f}i}=0.7$.
The time developments of the crack fronts are
shown in Fig.\ref{fig6} for $\Delta=1.120$, 1.129, and 1.141.
The roughness
exponents $\alpha$ of the front are $\alpha=$0.67, 0.50, and 0.35,
respectively.  The general trend that the front line is smoother for
larger $\Delta$ is the same with the case of model A, but the front
line are substantially rougher for the model B and there is no clear
transition to the smooth front as the model A.

As for the case of $\Delta=1.120$, we also analyzed the time
development of the crack front roughening using the finite size
scaling.  We measured the width of the front line $W$ as a function of 
the system width $L$ and the distance that the crack front travels
$X$, and fit the data in the form,
\begin{equation}
W(L,X) \sim L^\alpha f\left( {X\over L^z}\right),
\end{equation}
where $z$ is the dynamical exponent and $f(x)$ is a scaling
function,
\begin{equation}
f(x)\sim \left\{
\begin{array}{ll}
x^\beta & (x<<1) \\ \mbox{ const.} & (x>>1)
\end{array}\right.
\end{equation}
with $\beta$ being the growth exponent and the scaling relation
$z=\alpha/\beta$ holds.
The scaling plot is shown in Fig.\ref{fig7} with $\alpha=0.70$ and
$\beta=1.4$ for $L=$512, 256, 128, and 64.

\section{Discussion}

We have studied the stability of crack front numerically using a
simple 1-d mass-spring system, and found the straight crack front
propagation is not stable especially for slower crack speed.
This contrasts with the dynamical instability of the crack surface,
where the instability takes place for higher speed.\cite{FGMS91,ML93}

For the uniform model perturbed by the localized impurities, there
seems to be a rough-smooth transition at the certain speed beyond which
the crack front becomes smooth.
In the rough region at slower crack speed, the roughness exponent is
fairly constant around 0.4.

For the inhomogeneous model where the bond breaking threshold is
random, the front line also becomes smoother for the higher speed, but
the transition from rough to smooth front is gradual.  For the slowest
case we simulated, the obtained exponent $\alpha=0.70$ is to be
compared with 0.75 of the roughness exponent for the KPZ system with
the quenched noise, or with 0.633 of the directed percolation.

On the other hand, however, the growth exponent $\beta=1.4$ and the
dynamical exponent $z=0.5$, which are related to the dynamical
process, are quite different from those systems, where $\beta=0.61$
and $z=1.16$ for KPZ with the quenched randomness,\cite{KLT91} and
$\beta=0.633$ and $z=1$ for the directed
percolation\cite{TL92,BBCHSV92}.

The experiment which is closely related to the present work has been
done on the Plexiglas,\cite{SM97} and the exponent was found to be
around $0.55\pm 0.05$.
This is not very different from the one we obtained for the uniform
model, 0.4, but it should be noted that there are major differences
in the experimental situations from the present calculation;
1) the crack speed in the experiment is $10^{-7}\sim 10^{-5}m/s$ and
much slower than the sound speed,
2) there should be some long range interaction due to the three
dimensionality even though the experiment was done on the plate whose
thickness is much smaller than the width.

\acknowledgement

The work is partially supported by Grant-in-Aid for Scientific Research 
(c), the Ministry of Education, Science, Sports, and Culture (grant \#
09640468), and the Hong Kong Research Grants Council Grant No. 315/96P.



\newpage

\noindent{figure 1}

\newpage

\noindent{figure 2}

\newpage

\noindent{figure 3}
\vspace*{2cm}

\begin{figure}[htbp]
\begin{center}
  \epsfig{file=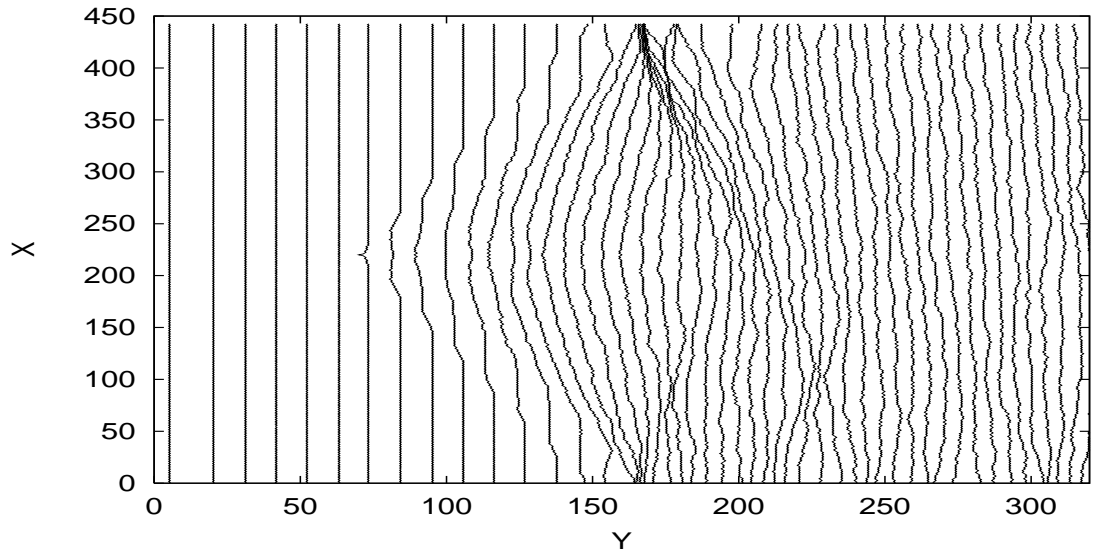,height=6cm,width=14cm,angle=-90}\par
\end{center}
\caption{
Time development of the crack front.
The steady solution is disturbed by a single impurity.
}
\end{figure}
\label{fig3}
\vspace{1cm}

\begin{figure}[htbp]
\begin{center}
 \psfig{file=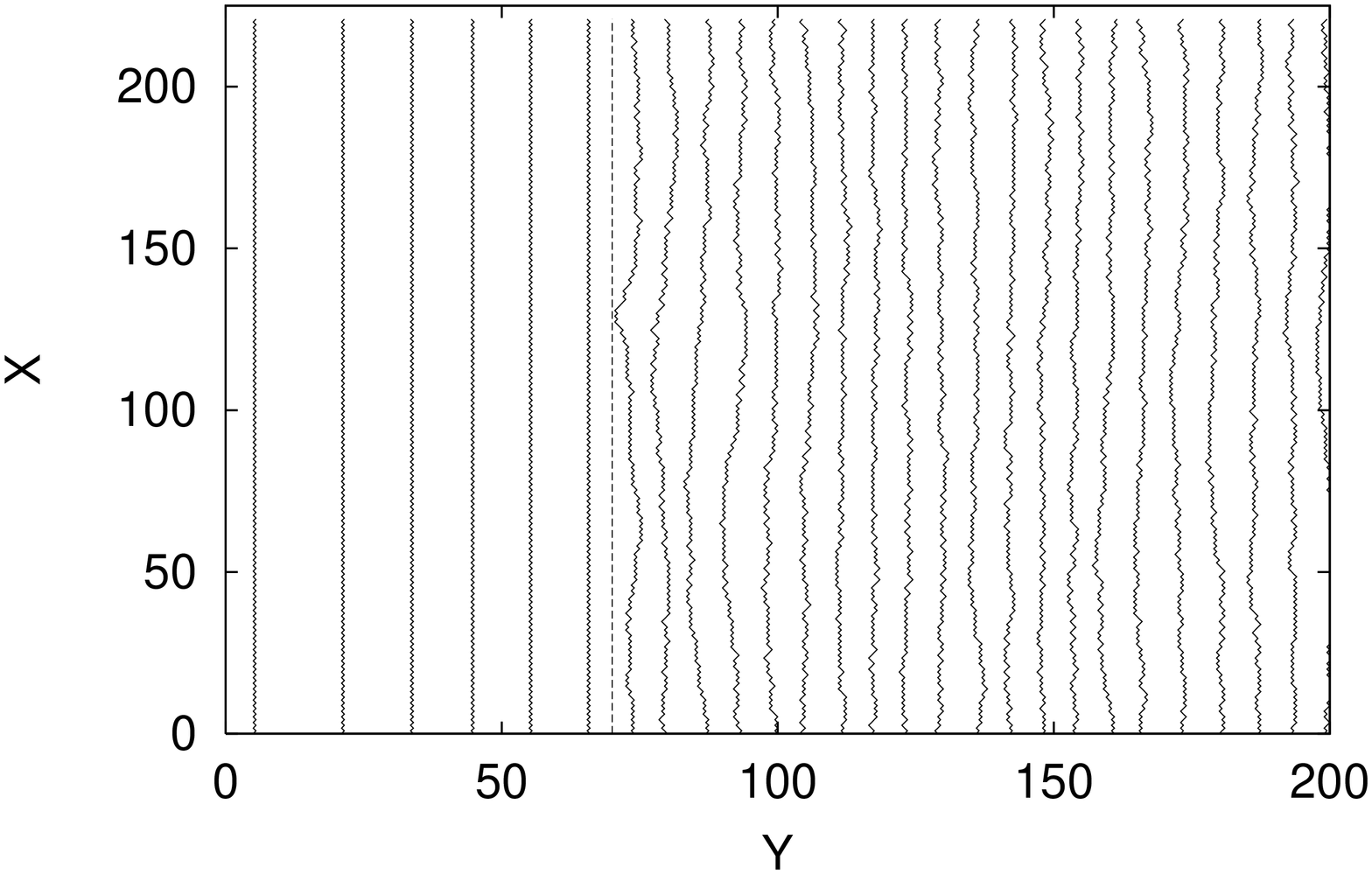,height=5.5cm,width=13cm,angle=-90} \vskip 1.cm \par
 \psfig{file=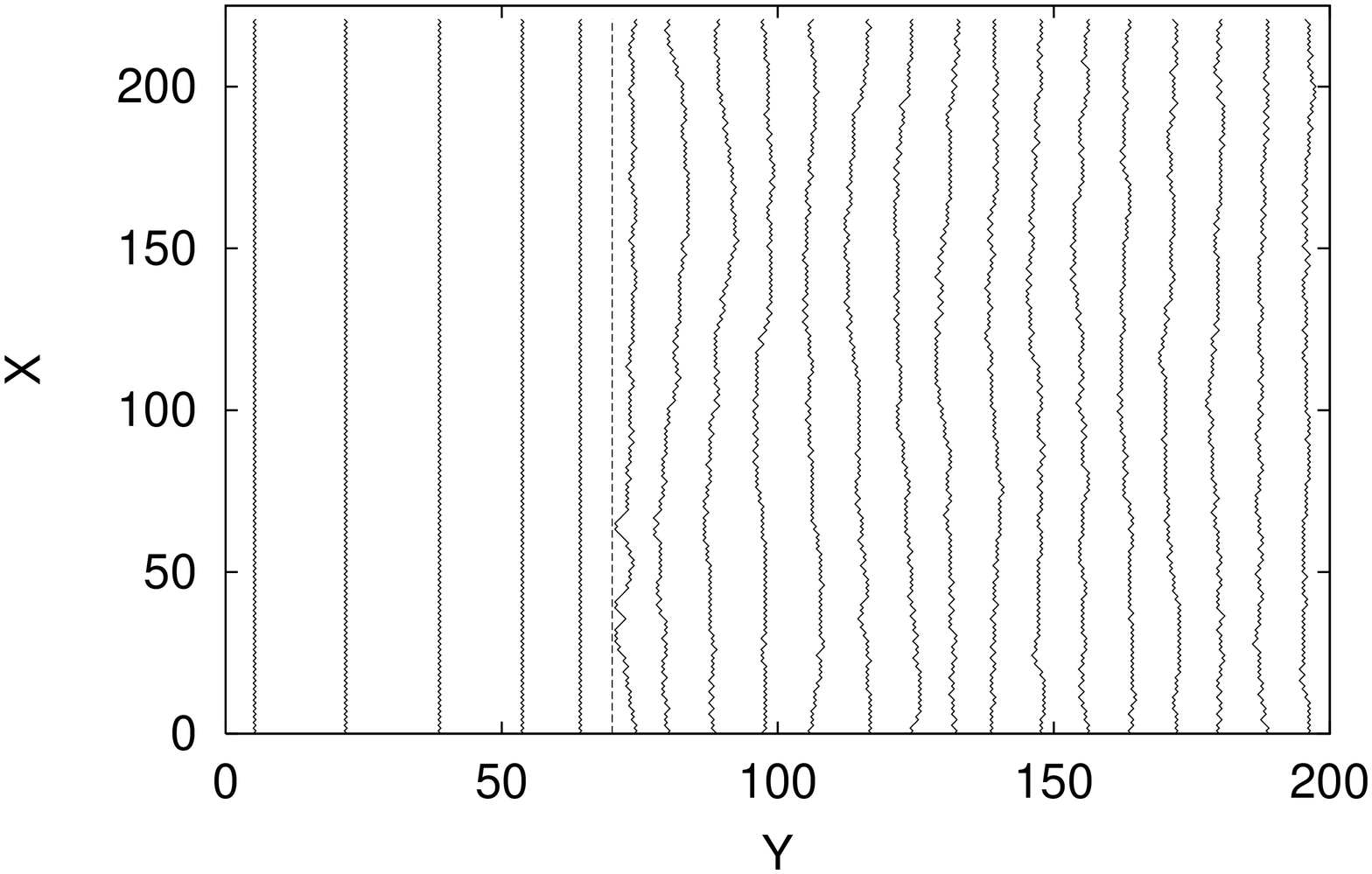,height=5.5cm,width=13cm,angle=-90} \vskip 1.cm \par
 \psfig{file=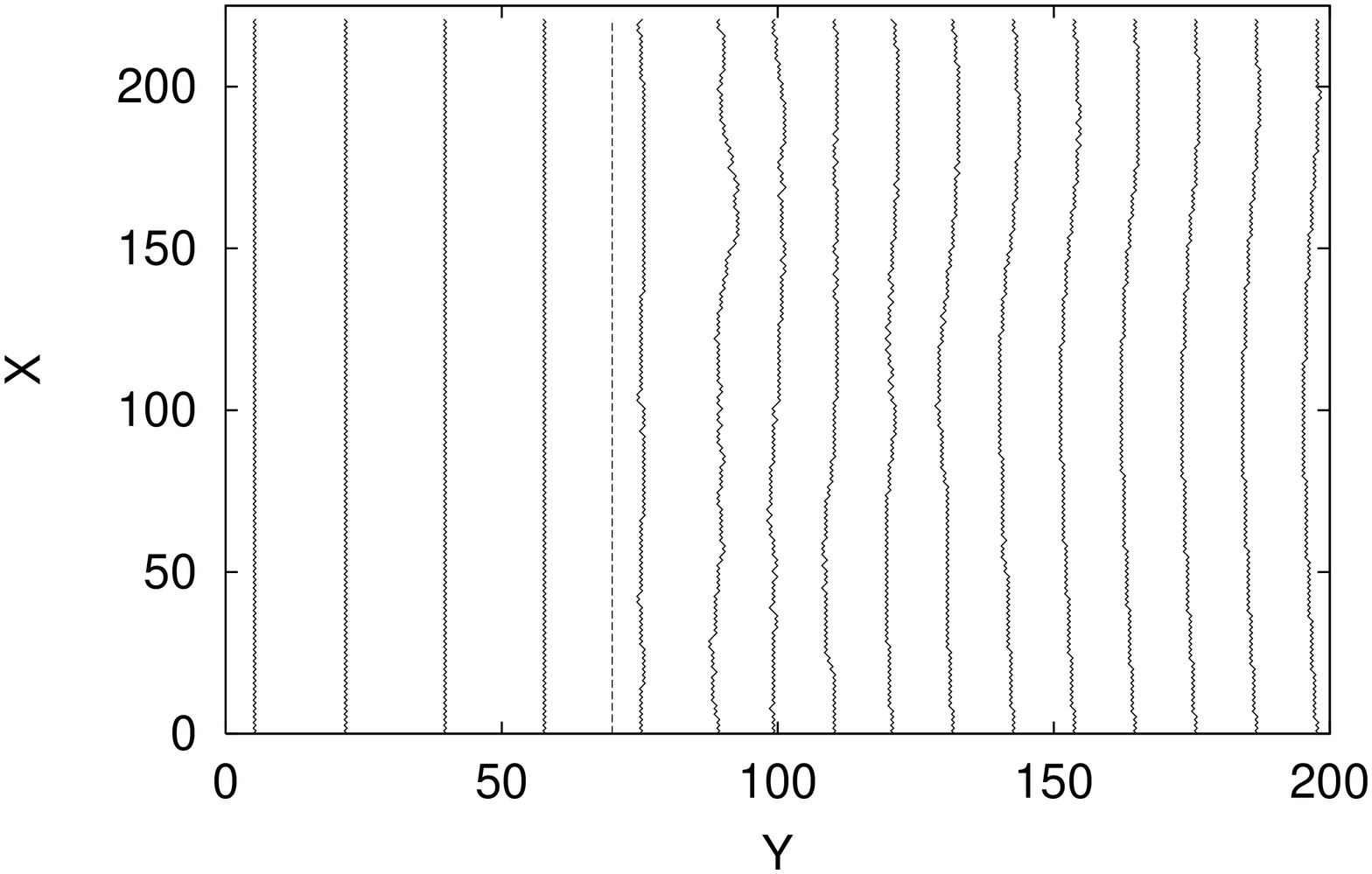,height=5.5cm,width=13cm,angle=-90}\par
\end{center}
\caption{
Time developments of the crack front.
The external displacements are $\Delta=1.049$ (top), 1.052 (middle), and
1.053 (bottom).
}
\label{fig4}
\end{figure}


\newpage

\vspace*{1cm}

\noindent{figure 5}
\vskip 1cm

\begin{figure}[htbp]
\begin{center}
  \leavevmode
  \psfig{file=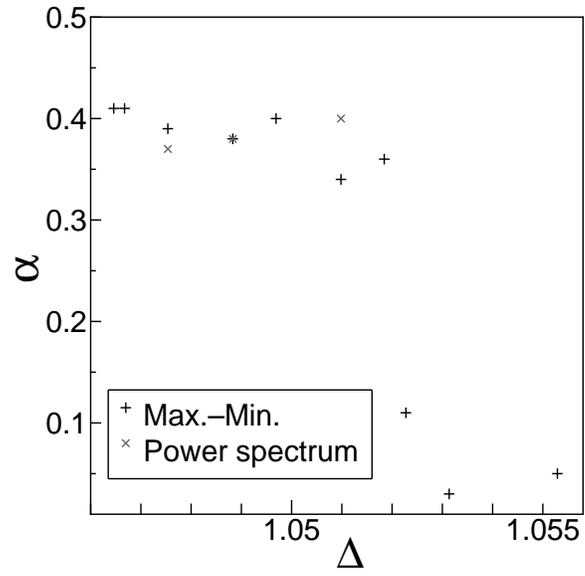,height=8cm,width=8cm}
\end{center}
\caption{
Roughness exponents $\alpha$ {\it v.s.} the external displacement
$\Delta$ for the model A.
}
\label{fig5}
\end{figure}


\begin{figure}[htbp]
\begin{center}
  \psfig{file=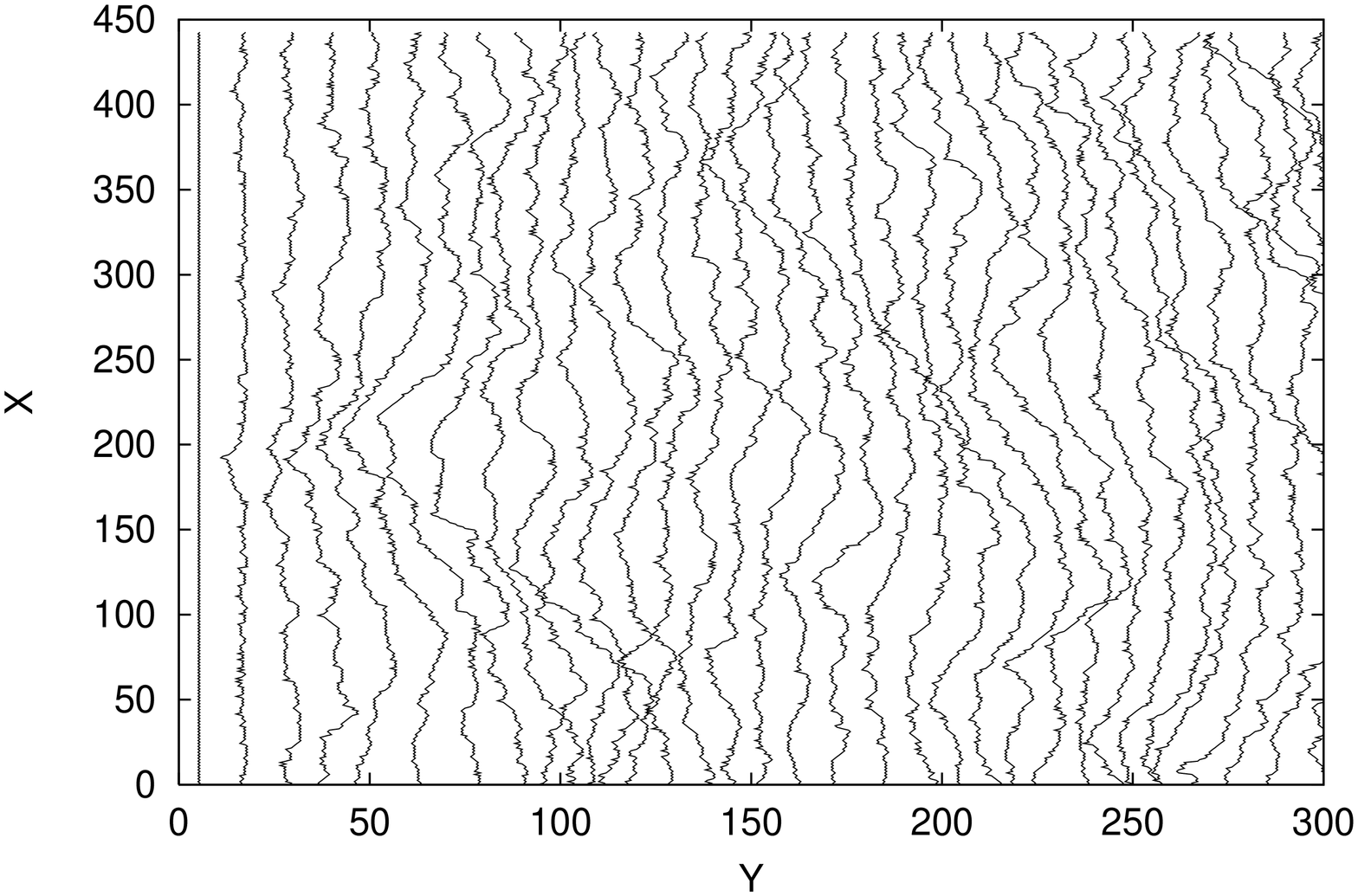,height=6.cm,width=13cm,angle=-90} \par
  \psfig{file=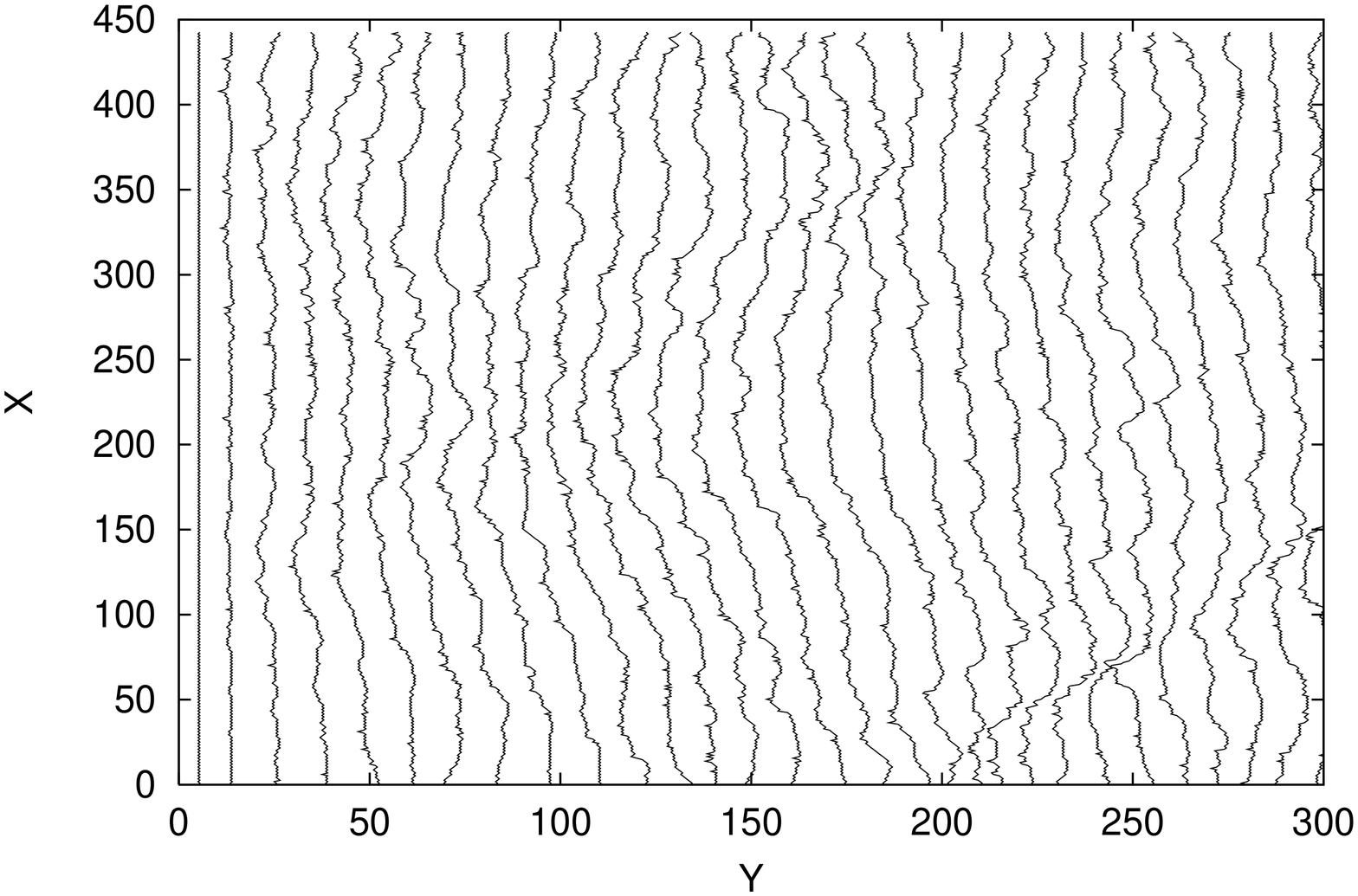,height=6.cm,width=13cm,angle=-90} \par
  \psfig{file=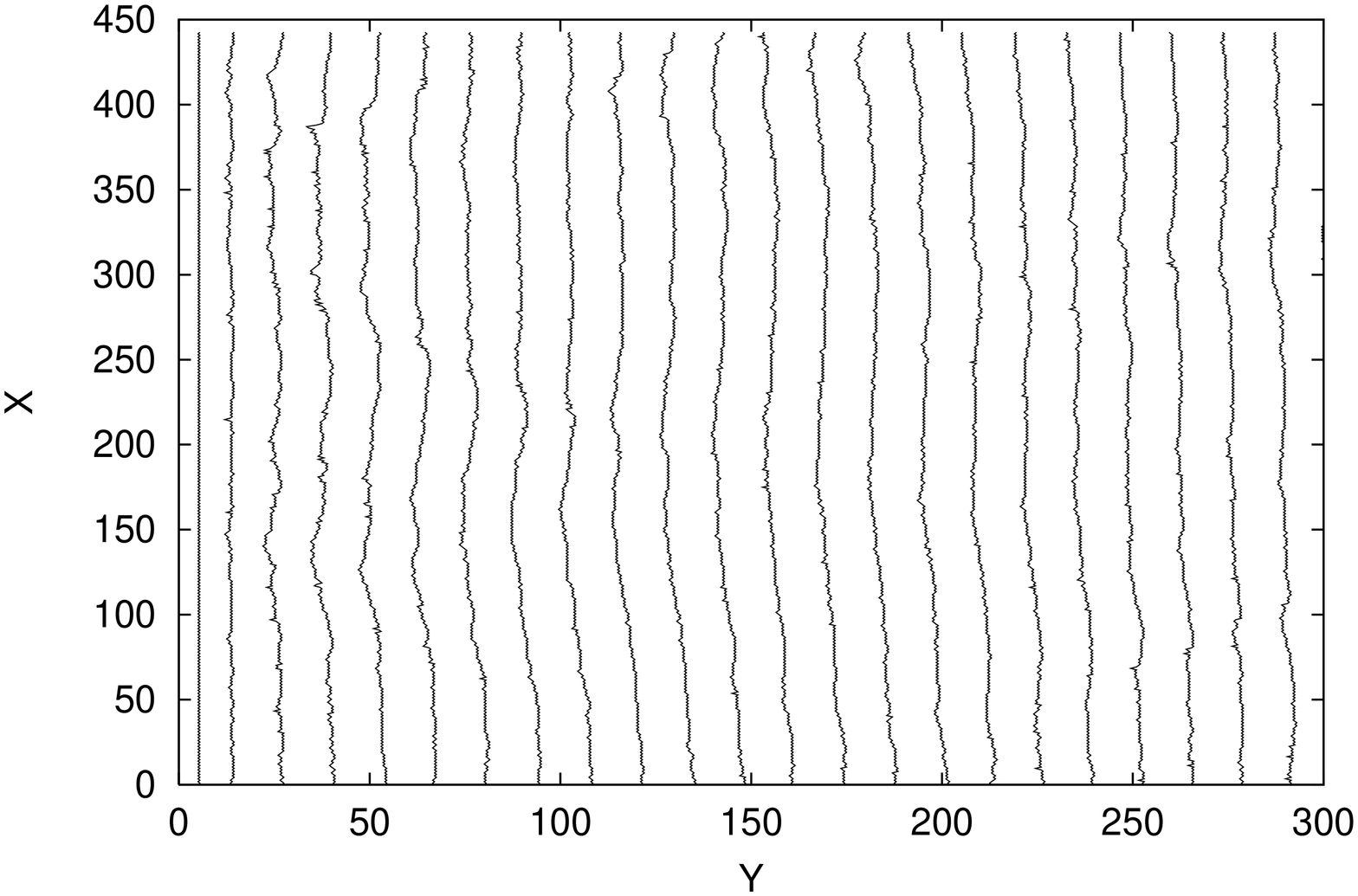,height=6.cm,width=13cm,angle=-90}\par
\end{center}
\caption{
Time development of the crack front for the model B.
The external displacements are $\Delta=$1.120 (top), 1.129 (middle), and
1.141 (bottom).
}
\label{fig6}
\end{figure}

\newpage

\vspace*{1cm}

\noindent{figure 7}
\vskip 1cm

\begin{figure}[htbp]
\begin{center}
  \leavevmode
  \psfig{file=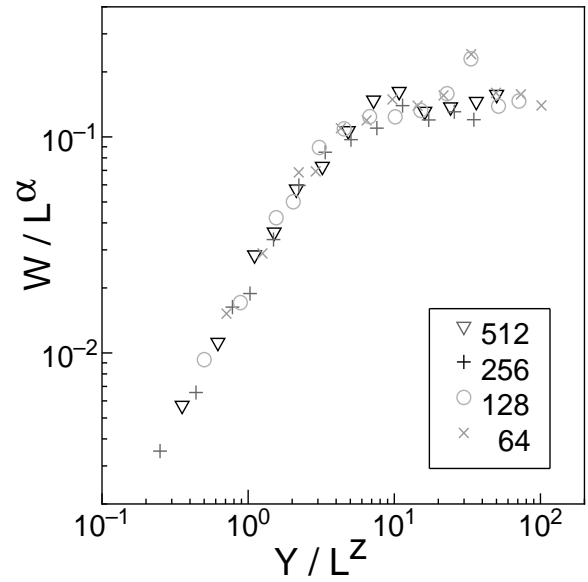,height=8cm,width=8cm}
\end{center}
\caption{
Scaling plot for the time development of the crack front in the model B
with $\Delta=1.120$.
The fitting parameters are $\alpha=0.70$ and $\beta=1.4$.
}
\label{fig7}
\end{figure}


\end{document}